\documentclass{article}

\usepackage[final,nonatbib]{neurips_2020}

\usepackage[utf8]{inputenc} % allow utf-8 input
\usepackage[T1]{fontenc}    % use 8-bit T1 fonts
\usepackage{hyperref}       % hyperlinks
\usepackage{url}            % simple URL typesetting
\usepackage{booktabs}       % professional-quality tables
\usepackage{amsfonts}       % blackboard math symbols
\usepackage{nicefrac}       % compact symbols for 1/2, etc.
\usepackage{microtype}      % microtypography
\usepackage{graphicx}
\usepackage{subfigure}
\usepackage{amsmath}

\title{Active Learning from Crowd in Document Screening}

\author{%
 Evgeny Krivosheev\\
  University of Trento, Italy\\
  \texttt{evgeny.krivosheev@unitn.it} \\
  \And
Burcu Sayin\\
  University of Trento, Italy\\
  \texttt{burcu.sayin@unitn.it} \\
  \And
    Alessandro Bozzon\\
  TU Delft, The Netherlands\\
  \texttt{a.bozzon@tudelft.nl} \\
  \And
   Zoltán Szlávik\\
 myTomorrows, The Netherlands\\
  \texttt{zoltan.szlavik@mytomorrows.com} \\
}

\begin{document}

\maketitle

\begin{abstract}
In this paper, we explore how to efficiently combine crowdsourcing and machine intelligence for the problem of document screening, where we need to screen documents with a set of machine-learning filters. Specifically, we focus on building a set of machine learning classifiers that evaluate documents, and then screen them efficiently. It is a challenging task since the budget is limited and there are countless number of ways to spend the given budget on the problem. We propose a multi-label active learning screening specific sampling technique -\textit{objective-aware sampling}- for querying unlabelled documents for annotating. Our algorithm takes a decision on which machine filter need more training data and how to choose unlabeled items to annotate in order to minimize the risk of \textit{overall classification} errors rather than minimizing a single filter error. 
We demonstrate that objective-aware sampling significantly outperforms the state of the art active learning sampling strategies.
% on multi-filter classification problems.

\end{abstract}

\section{Introduction}
\begin{figure}[tbh]
    \centering
    \includegraphics[width=0.6\textwidth]{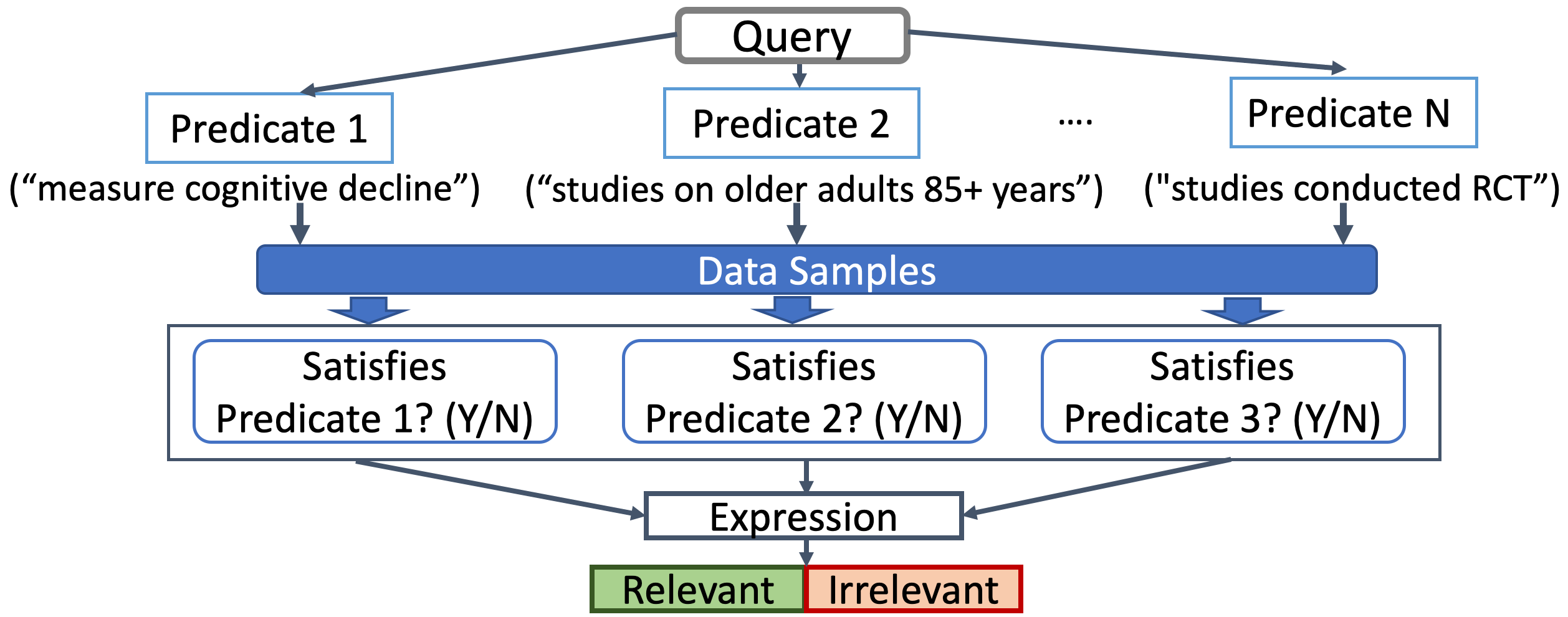}
    \caption{Predicate-based screening process of documents}
    \label{fig:screening}
\end{figure}

In this paper, we tackle the problem of screening a finite pool of documents, where the aim is to retrieve relevant documents \textit{satisfying a given set of predicates} that can be verified by human or machines (Fig. \ref{fig:screening}). In this context, if a document does not satisfy at least one predicate, it is treated to be irrelevant. A predicate represents a property, a unit of meaning, given in natural language (e.g., "find papers that measure cognitive decline"). By this means a predicate might be interpreted in a variety of ways in text, so making keywords-based search hard to reach high recall while keeping a decent level of precision \cite{ZhangAbualsaud2018}. We interpret the screening problem as high recall problem, i.e., the aim is to retrieve all relevant documents maximizing precision. %\textbf{we assume predicates and candidate documents are given}
% Since predicates can be interpreted in a variety of ways, it makes the problem of document screening very challenging especially when there is a little training data. 
The screening finds application in many domains, such as i) in systematic literature reviews \cite{Wallace2017};
% -SLRs- ( - where in-scope papers are those that match a set of inclusive predicates \cite{Wallace2017}, e.g., \textit{papers measure cognitive decline (predicate 1) AND papers studying older adults (predicate 2)};
ii) database querying - where items filtered  in/out based on predicates \cite{Parameswaran2014}; iii) hotel search - where the hotels retrieve  based upon filters of interest \cite{Lan_dynamicfilter_hcomp17}. Consequently,  the document screening is an instance of finite pool binary classification problems  \cite{Nguyen2015}, where we need to classify a finite set of objects minimizing cost.
% As an instance of the problem, we choose the screening phase of SLRs what makes the problem rather challenging since each review is different and has a unique set of predicates (in SLRs called inclusion criteria). Typically, authors of an SLR retrieve a candidate pool of documents executing a keywords-based query on a database such as Scopus. To avoid missing papers, the query tends to be inclusive, which means that it returns hundreds or thousands of results (typically 4-5k on average \cite{Nguyen2015}) that are later manually \textit{screened} by researchers based on predefined predicates. For example, researchers might look for papers that describe all of the following predicates at the same time: 1) "include papers that study older adults 85+ years", 2) "include papers conducted randomized controlled trial", 3) " include papers about behavioral intervention". Therefore, here we have the conjunctive query of three inclusive predicates.
A bottleneck of the screening process is the predicate evaluation, i.e.,  identifying which of the given predicates are satisfied in a current document. For example, in literature reviews, authors validate predicates, however, this is time-consuming, exhaustive, and very expensive \cite{coch_handbook2011,Krivosheev_hcomp17,KrivosheevWWW2018,Sampson2008}. 

An effective technique to solve screening problems is \textit{crowdsourcing} where the crowd can solve even complex screening tasks with high accuracy and lower cost compared to expert screening \cite{Law_hearth_cscw18,Krivosheev_hcomp17,KrivosheevWWW2018,Mortensen2016crowd,Parameswaran2014,ParameswaranACM2012,Ranard_harnessing_2014,SunHCOMP2016,Wallace2017}. However, achieving a good performance in crowd-based screening requires a deep understanding of how to design tasks and model their complexity \cite{GadirajuHT17,quinn_17_confusing,QaroutCEUR18,yang2016modeling}, how to test and filter workers \cite{Bragg2016}, how to aggregate results into a classification decision~\cite{DawidSkene_Confusion,Dong2013,HanWSDM20,karger2011iterative,KrivosheevWWW2018,Li_error_13,LiuWang_truelabel,liu2013scoring,Ok_belief_13,whitehill2009whose,Zhou_spectral_13}, or how to improve worker engagement \cite{HanWSDM19,HanIEEE19,QuiCHI2020}. 

Machine learning (ML) algorithms have also made a very impressive progress in solving complex screening tasks. However, obtaining a sufficiently large set of training data is still a key bottleneck for accurate ML classifiers. Active learning (AL) \cite{Cohn96} accelerates this process by minimizing the size of training data that is required to train better classifiers via selecting the most informative instances for annotation. The effectiveness of AL have been proven in many domains (see the surveys \cite{Aggarwal_2014_AL,Settles10ALSurvey}), but most of the work considers single-label cases while multi-label AL problems have been far less investigated. The challenge in applying AL to a multi-label classification problem is that the algorithm should measure the unified informativeness of each unlabeled item across all labels. The state of the art multi-label AL strategies follow an object-wise (global) labeling, where the AL algorithm first finds the relevance scores (i.e. confidence-based or disagreement-based scores) of $<item,label>$ pairs, and then aggregates these scores to find the informativeness of items \cite{cherman2016active,Huang2015IJCAI,Vasisht2014,Wu2014MLAL, Ye2015}. However, it may ignore the interaction between labels \cite{ReyesMLAL2018}.

\paragraph{Original contribution.} We investigate how to efficiently combine crowdsourcing and ML for item screening. It is a challenging task since the budget is limited and there are countless number of ways to spend it on the problem. We propose a multi-label AL screening specific sampling technique for querying unlabelled items for annotating. Our algorithm takes a decision how to choose unlabeled data to annotate by crowd workers in order to maximize the performance of a screening task. Unlike existing multi-label AL approaches that rely on global labeling, we choose the local labeling method, where for each label (predicate in our case) we determine the relevancy to each item.

\section{Approach}

\subsection{Problem Statement}
We model the problem as the input tuple $(P, UI, B, E, AL)$, where
 $P = \{p_1, .., p_n\}$ is the set of inclusive filters expressed in natural language, e.g., \textit{P = ("papers study older adults",  "papers study mental disorders")}, $UI$ is the set of unlabeled items (documents) that needed to be labeled as either "IN" or "OUT" of scope,
% \item $LI^{0}$ is an initial little set of training data items for machine-learning binary classifiers $C = \{c_1, .., c_n\}$. Each classifier is responsible for a filter $f$ from $F$ and outputs the probability of applying $f$ on document $i$.
 $B$ is an available budget for collecting training data for classifiers $A$ (we do not have training data in the beginning), $AL$ is an active learning strategy, and $E$ is a conjunctive expression of predicates.
The output is the tuple $(LI, A)$, where $LI$ is the labeled items,
$A$ is the set of trained binary ML classifiers that evaluate if an item describes the predicates.
% Worth noting, classifiers $A$ can be utilized for a new screening task that share the same predicates, and thereby we will have a subset of already trained classifiers. 
The screening operates on each unlabeled item estimating how much predicates $P$ match the current document $i$. If at least one of the predicates does not match, the document is considered as irrelevant. We can compute the probability that a document $i$ should be excluded: $Prob(i \in OUT) = 1 - \prod_{p \in P} Prob(i_p \in IN)$.
% Therefore, our aim is to annotate a subset of unlabeled documents $UI$ via $AL$ and crowd workers within budget $B$ and then to classify all $UI$ with trained classifiers $A$. 
In this work, we propose a new $AL$ strategy for multi-predicate classification and evaluate it under different settings.
We observe that rather than having one ML algorithm for the overall screening problem, it is often convenient to have a separate classifier $A^p_{ml}$ for each predicate $p$.
There are several reasons for this: besides being a requirement of some crowd-machine algorithms~\cite{Nguyen2015, KrivosheevCSCW2018} from budget optimisation to facilitating their reuse \cite{cherman2016active, ramirez2018crowdrev}. 

\subsection{Objective-Aware Sampling}\label{sec:alscreening}
We now discuss how ML classifiers for each predicate can be trained, given an ML algorithm (e.g., SVM).
The most common AL method is \textit{uncertainty sampling}, that selects items closest to a classifier's decision boundary \cite{Aggarwal_2014_AL}. We introduce here an \textit{objective-aware sampling} technique particularly suited for screening problems and that, as we will show, can outperform uncertainty sampling. We motivate the need for -- and the rationale of -- \textit{objective-aware sampling} with an example.
We want to train two binary classifiers ($A^1_{ml}, A^2_{ml}$) to evaluate predicates $p_1, p_2$ respectively. As we iterate through the training set to learn the model, assume we reach the situation shown in Table \ref{tab:m_example}. In AL fashion we then need to choose the next \textit{(item, predicate)} to label.

\begin{table}[h]
  \caption{Objective-aware sampling queries items for a classifier based on overall screening uncertainty. \\}
  \label{tab:m_example}
  \centering
  \scalebox{0.75}{%
  \begin{tabular}{cccl}
    \toprule
    Item ID& 
        \begin{tabular}[x]{@{}c@{}}
            $Prob(i_{p_1} \in IN)$\\estimated by $A^1_{ml}$\end{tabular}
        &\begin{tabular}[x]{@{}c@{}}
            $Prob(i_{p_2} \in IN)$\\estimated by $A^2_{ml}$\end{tabular}
        &\begin{tabular}[x]{@{}c@{}}True class\\(hidden)\end{tabular}\\
    \midrule
    1 & 0.99 & 0.98& IN\\
    2 & 0.51 & 0.01 & OUT\\
    \textbf{3} & \textbf{0.51} & \textbf{0.99} & \textbf{OUT}\\
    4 & 0.03 & 0.01 & OUT\\
  \bottomrule
\end{tabular}}
\end{table}

In this context, if we want to improve $A_{ml}^1$ independently from the overall screening objective, then we adopt uncertainty sampling as per the literature, choosing either item 2 or 3. 
However, the classification errors of $A_{ml}^1$ might have a bigger impact if predicate $p_2$ applies to the item (that is, if the item is IN scope according to $p_2$), because then the final decision rests on $p_1$ and therefore $A_{ml}^1$. 
In the table above, the risk of overall classification errors for item 2 is low as $A_{ml}^2$ will filter out the item anyways, regardless of the opinion of $A_{ml}^1$. This is not the case for item 3 which is therefore a more promising case to pick for labeling.
In other words, denoting with $P^p$ the estimation by the ML classifier $A_{ml}^p$ trained on predicate $p$, 
 we choose the item for which the chance of making the \textit{False Negative} error is the highest.
In summary, we apply a combination of uncertainty and certainty sampling, but we look at the screening accuracy rather then only at the current classifier:

\begin{equation}
	\begin{aligned}
		score(i, \bar{p}) = (1-Prob(\hat{y} | i, A_{ml}^{\bar{p}})) \cdot \prod_{p \in \{P-\bar{p}\}} Prob(i_p \in IN)
	\end{aligned}
	\label{formula:score}
\end{equation}
where $\bar{p}$ is the predicate for which we want to improve the ML classification, $\hat{y} = argmax_y P(y|i)$ is the class label with the highest posterior probability , $\{P-\bar{p}\}$ the set of predicates except one we want to improve on.
We then chose top-k (item, predicate) with the highest scores for labeling. 
Notice that objective-aware sampling applies when classifiers are trained for each predicate. 

\section{Experiments and Results}

\textbf{Methodology.} We experiment objective-aware strategy with the goal of understanding whether there is an improvement in screening accuracy where we fix a budget that can be spent on crowd or expert annotations. Our baseline AL strategies include uncertainty sampling and random sampling. 

To do this, we identify datasets with different characteristics in terms of accuracy that machine or crowd achieve, and in terms of predicate selectivity. The \texttt{Amazon Reviews} dataset~\footnote{\scriptsize{https://github.com/TrentoCrowdAI/crowdsourced-datasets}} contains 5000 reviews on products, with information about the binary sentiment expressed in the reviews and whether a review was written on a book. In this screening task, we have two predicates: the \texttt{Books} predicate (with selectivity of  0.61) and the \texttt{Negative review} (with selectivity of 0.10), only $5\%$ of reviews are therefore considered relevant. The rationale for this dataset is to have a scenario where the task is relatively easy for the crowd with the average crowd accuracy of 0.94.
\texttt{Systematic Literature Review} (SLR) dataset~\cite{KrivosheevWWW2018} mimics the screening phase of the SLR process, where the abstracts returned from a keywords-based search need to be screened by researchers.
This dataset contains 825 items for two predicates: \texttt{Does the experiment targeting and involving "older adults"?}  (with selectivity of 0.58, crowd accuracy is 0.8) and \texttt{Does the paper describe an intervention study?} (with selectivity of 0.20, crowd accuracy is 0.6), and $17\%$ of relevant items. Unlike the Amazon dataset, this task is hard for the crowd and even harder for ML classifiers.
Each of the datasets was partially labeled by crowd workers (e.g., 1k reviews in the Amazon dataset) which allows us to estimate the accuracy distribution of workers on predicates and to simulate crowd annotations with the real answer probability distributions of the datasets and thus to analyze the proposed strategy under different conditions.

For each predicate, we trained an SVM classifier with TF-IDF features and ``balanced" class weights. We use F1 score to measure the classification accuracy and report its average value over 10 runs. The source code and experiment results (including additional metrics) are available online~\footnote{\scriptsize{https://github.com/Evgeneus/Active-Hybrid-Classificatoin\_MultiPredicate}}

\begin{figure}[t]
    \centering
    \subfigure[]{\includegraphics[width=0.32\textwidth]{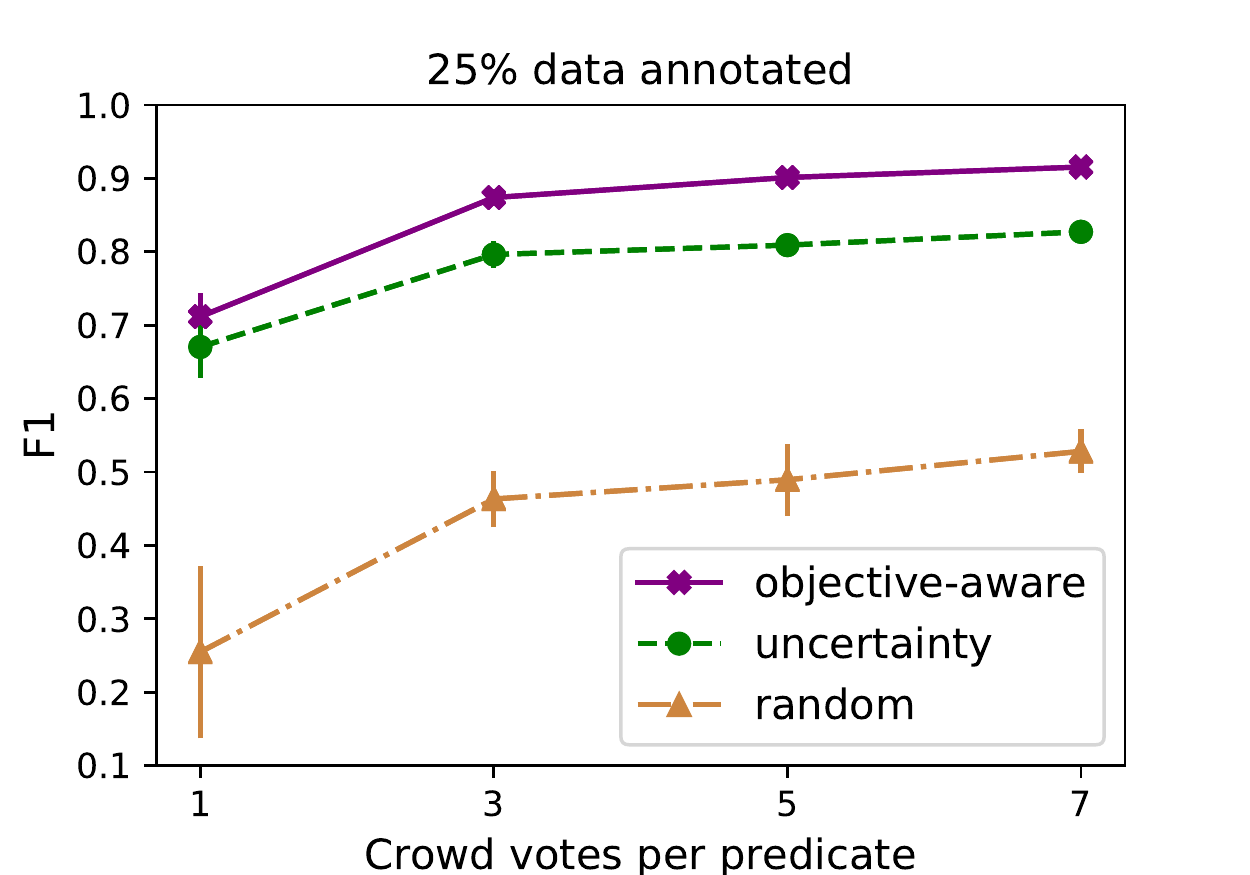}} 
    \subfigure[]{\includegraphics[width=0.32\textwidth]{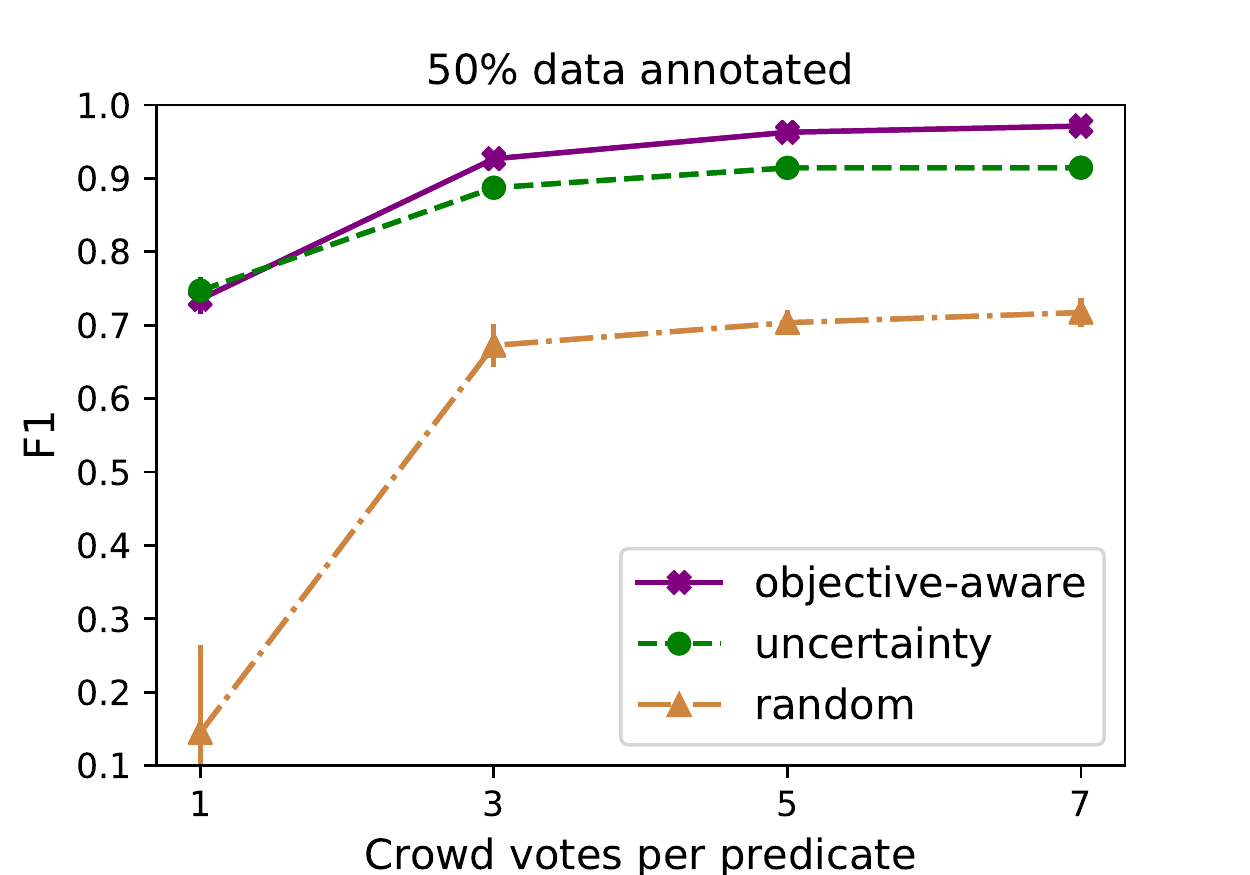}} 
    \subfigure[]{\includegraphics[width=0.32\textwidth]{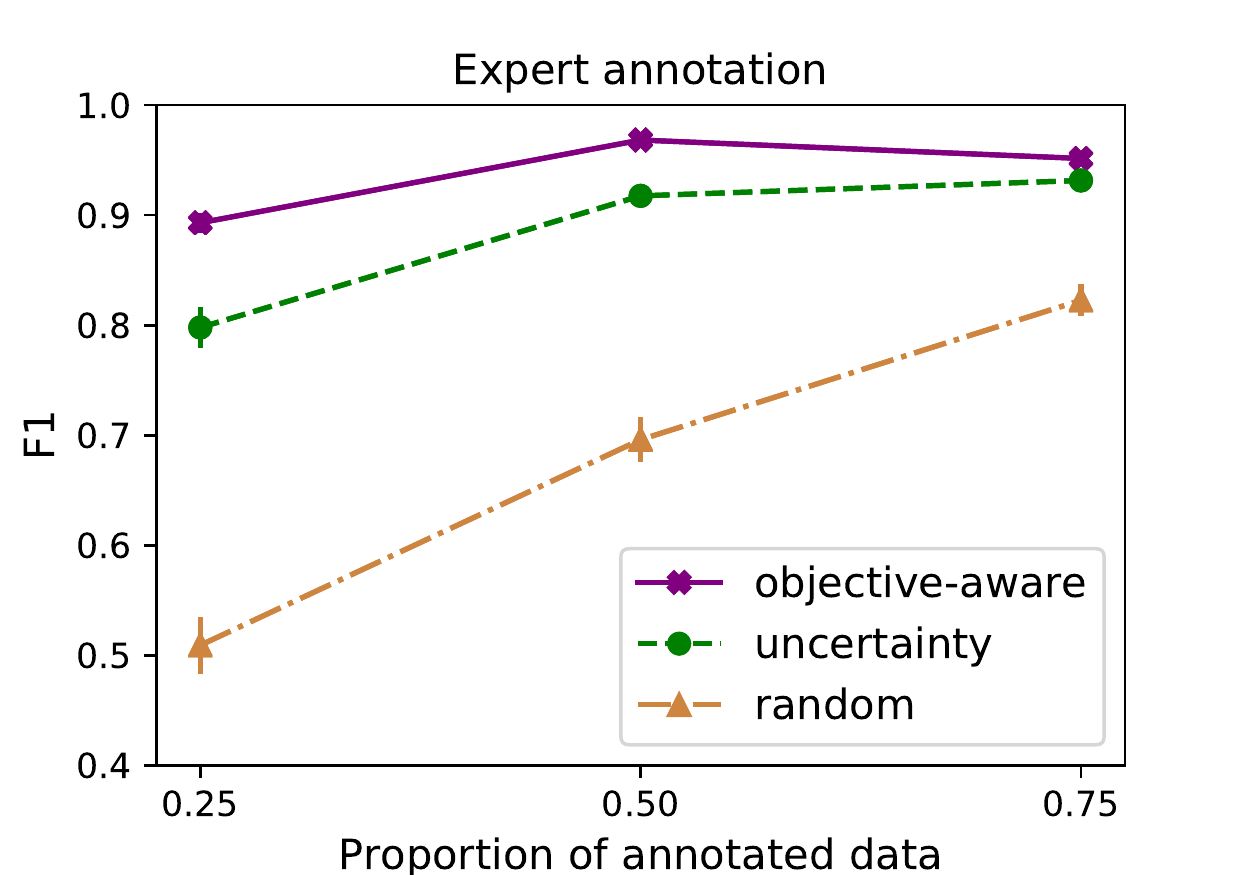}}\vspace{-1.1\baselineskip}
    \subfigure[]{\includegraphics[width=0.32\textwidth]{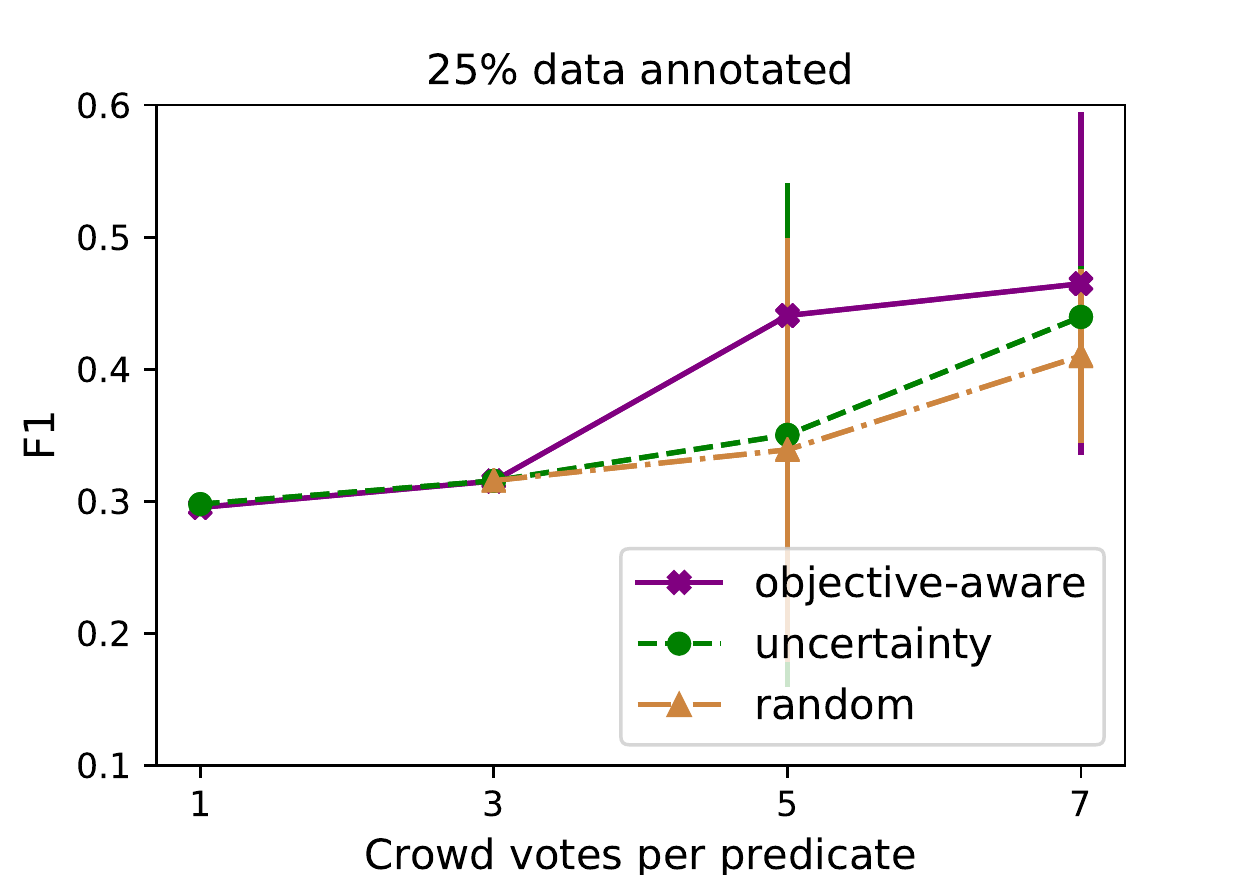}}
    \subfigure[]{\includegraphics[width=0.32\textwidth]{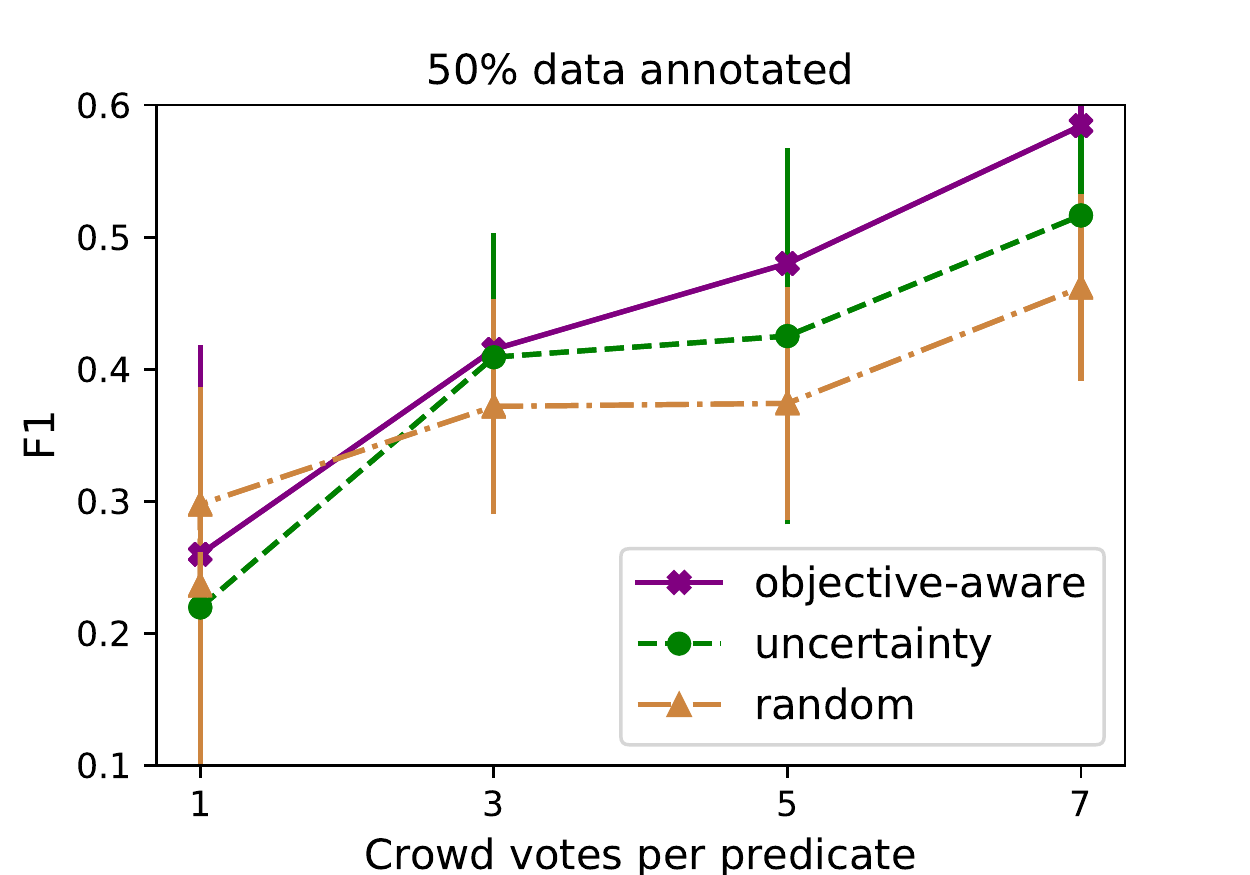}}
    \subfigure[]{\includegraphics[width=0.32\textwidth]{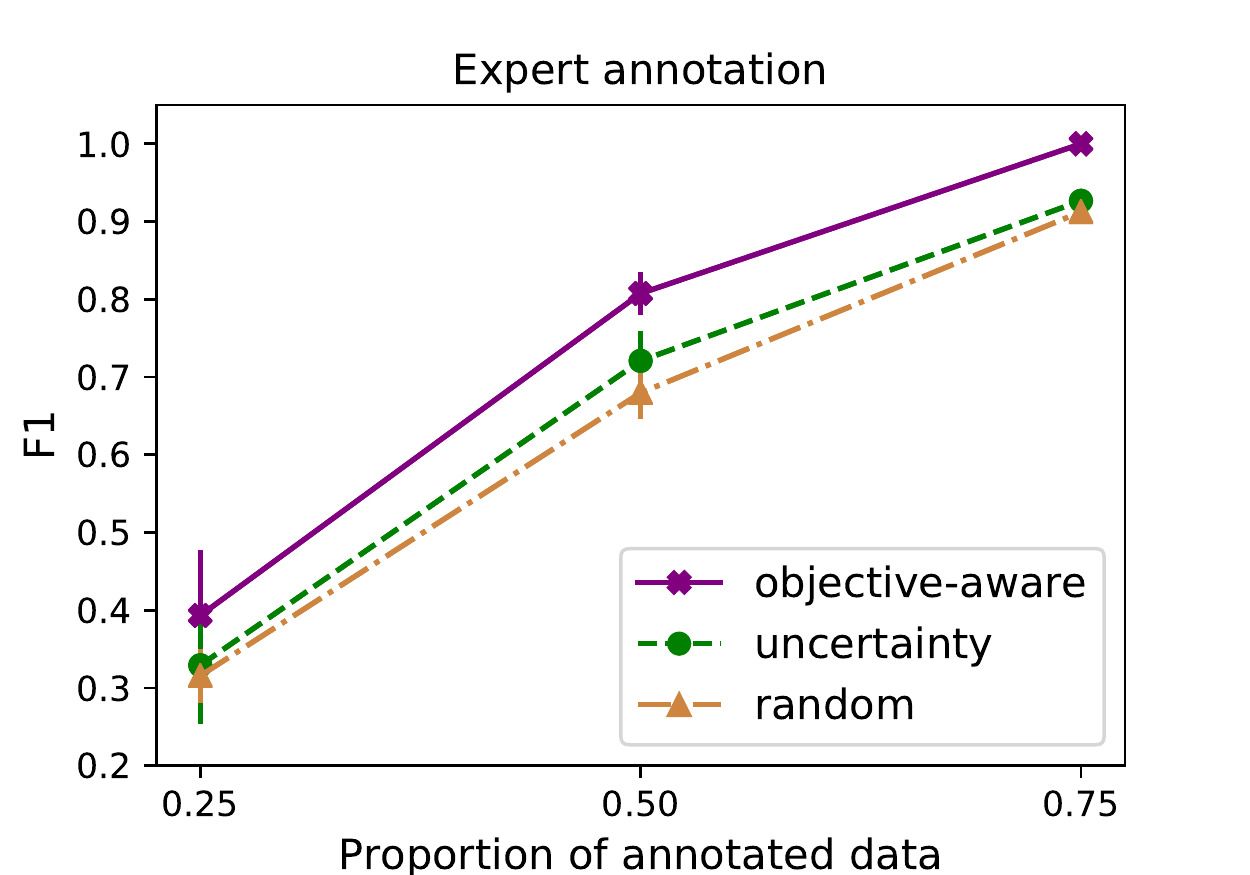}}
    \caption{Results for Amazon (a, b, c) and SLR (d, e, f) datasets}
    \label{fig:res}
\end{figure}

\textbf{Results.} To analyze the proposed AL strategy, we run experiments over the available datasets varying the budget, the proportion of annotated data, and the quality of annotations. Figure~\ref{fig:res} depicts the F1 screening score for $25\%$ and $50\%$ of annotated data samples. Along X-axis, we increase the number of crowd votes requested per (item, predicate) during the annotation process that followed by Majority Voting aggregation. Intuitively, the more crowd votes were queried per item the better accuracy of labels, and the more budget we need to spend considering a fixed proportion of annotated data. We observe for the Amazon dataset, where crowd accuracy is high, the objective-aware sampling always outperforms the baselines. Notably, with more accurate annotations (more crowd votes per predicate) our approach makes the accuracy gap between baselines even bigger thus outperforming uncertainty sampling by 7.8 and 8.8 points in F1 when collecting 3 and 7 votes/predicate respectively. SLR dataset is very difficult for both crowd and machines, and it is crucial to obtain high quality annotations. With less than 3 votes per predicate, the labels become highly inaccurate leading to poor screening accuracy. When we request 3 or more votes per predicate the results become much better and, as in the Amazon dataset, our objective-aware sampling strategy superiors both uncertainty and random sampling techniques by a big margin (Figure~\ref{fig:res} d, c).

We further examine how objective-aware sampling behaves in case of noise-free annotations obtained by experts.
Figure~\ref{fig:res} c,f shows the resulting F1 scores for different proportions of labeled data. We can see that our approach outperforms the baselines for both Amazon and SLR datasets on all the settings. Thus for $50\%$ of annotated data objective-aware sampling beats uncertainty sampling by 5.1 and 8.7 in F1 points for Amazon and SLR datasets respectively.

\section{Conclusion}
In this paper, we proposed and evaluated the objective-aware active learning strategy designed for screening classification and selecting efficiently item, predicate for annotating based on the overall classification objective. We demonstrated that objective-aware sampling outperforms uncertainty and random AL techniques under different conditions.
We further aim to examine more screening datasets, extend this study to other classes of screening problems and hybrid crowd-machine algorithms.

\bibliographystyle{plain}
\bibliography{main}

\end{document}